\documentclass[preprint,12pt,review]{elsarticle}

\usepackage{amssymb}
\usepackage{amsmath}

\journal{Physics Letters B}

\begin{document}

\begin{frontmatter}



\title{Neutrino photoproduction on pseudo Nambu--Goldstone bosons}


\author{I. Alikhanov\corref{cor1}}

\cortext[cor1]{{\it Email address: {\tt ialspbu@gmail.com}}}

\address{Institute for Nuclear Research of the Russian Academy of Sciences,
60-th October Anniversary pr. 7a, Moscow 117312, Russia}

\begin{abstract}
Production of single neutrinos as well as neutrino--antineutrino pairs by photons interacting with pseudo Nambu--Goldstone bosons is studied within the Standard Model. The corresponding cross sections are found analytically. The energy loss due to neutrino emission in a thermal plasma of photons and pions is calculated. It is shown that the obtained neutrino emissivity may be significantly enhanced in hot and dense matter due to in-medium modification of the pion decay constant. Phenomenological consequences for ultrarelativistic heavy-ion collisions and astrophysics are discussed. 
\end{abstract}


\end{frontmatter}

\section{Introduction}
Nambu--Goldstone bosons (often referred to as Goldstone bosons) appear necessarily in quantum field theories with spontaneously broken global continuous symmetries~\cite{nambu,goldstone,salam}. The bosons remain massless provided the symmetries are exact, acquiring masses only in the case of approximate symmetries. In the latter case they are called  pseudo Nambu--Goldstone bosons (pNGB). 

In nature, pNGB manifest themselves as the lightest pseudoscalar mesons from the SU(3) flavor octet~--~the pions~\cite{goldstone,jona1,jona2}. This happens due to the spontaneous chiral symmetry braking in quantum chromodynamics. The kaons may also be identified as pNGB~\cite{nussinov}.

The PNGb modes appearing in dense matter formed inside astrophysical objects, such as core collapse supernovae and compact stars, could make a dramatic impact on their thermal evolution. 
In 1965 Bahcall and Wolf demonstrated that a neutron star containing free pions in its interiors would cool much faster through neutrino emission in comparison with the conventional mechanisms -- the modified Urca and bremsstrahlung neutrino processes~\cite{wolf}.  Since then, the role of PNGb modes for compact star cooling attracts much attention~\cite{maxwell,umeda,pion_cond1,pion_cond2,pion_cond3,pion_cond4}.

The extreme conditions can also be fabricated in ultrarelativistic heavy-ion collisions. 

In this Letter production of neutrinos in interactions of photons with PNGb is studied within the Standard Model. Specifically, the neutrino emissivities through the processes $\gamma+\pi^0\rightarrow\nu+\bar\nu$, $\gamma+\pi^+\rightarrow e^++\nu_e$ and $\gamma+\pi^+\rightarrow\mu^++\nu_{\mu}$ are calculated. 

\section{Neutrino--antineutrino pair photoproduction on neutral pNGB\label{sec0}}

\begin{figure}
\centering
\includegraphics{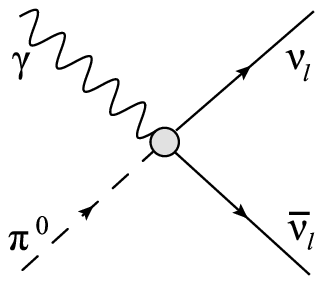}
\caption{}
\label{fig1}
\end{figure}

The Standard Model accommodates the following reaction:

\begin{equation}
\gamma+\pi^0\rightarrow\nu_l+\bar\nu_{l},
\label{fi0}
\end{equation}

represented by  the Feynman diagram in Fig.~\ref{fig1} ($l=e, \mu, \tau$). Processes related to~(\ref{fi0}) by crossing have been considered in~\cite{khlopov1,khlopov2,Harvey2007,plb_pion}.

The corresponding matrix element is~\cite{khlopov1,khlopov2,plb_pion}

\begin{equation}
{\cal M}=-\frac{eG_{F}}{\sqrt{2}m_{\pi}}F_{V}\varepsilon_{\mu }
\bar{u}(p'_{\nu})\gamma _{\alpha }(1-\gamma
_{5})v(p_{\nu})
\epsilon ^{\mu \alpha \beta
\lambda }q_{\beta }p_{\pi \lambda }. \label{ampl_2}
\end{equation}

Here $e$ is the elementary electric charge, $G_F$ is the Fermi coupling constant, $\varepsilon_{\mu}$ denotes the photon polarization vector,
$p_{\pi}$, $p_{\nu}$, $p'_{\nu}$, and $q$
are the four-momenta of $\pi^{0}$, the final neutrinos and $\gamma$, respectively, $F_{V}$ is
the pion vector form factor. 

Squaring (\ref{ampl_2}) yields

\begin{equation}
\sum_{\text{spins}}|{\cal M}|^2=\frac{\alpha\pi G_F^2}{m_{\pi}^2}|F_V|^2s\left(t^2+u^2\right),
\end{equation}

where $\alpha$ is the fine structure constant, $s=(p_{\pi}+q)^2$, $t=(p_{\pi}-p_{\bar\nu})^2$ and $u=(p_{\pi}-p_{\nu})^2$ are the Mandelstam variables.

After the standard algebra one obtains the cross section of~(\ref{fi0}) for each neutrino flavor:

\begin{equation}
\sigma_{\pi}=\frac{\alpha G^2_{F}}{24m^2_\pi}|F_{V}|^2s^2\left(1-\frac{m^2_{\pi}}{s}\right). \label{cross_sect_tot}
\end{equation}

Details of similar calculations can be found in~\cite{plb_pion}. Note that though the formfactor depends, in general, on the momentum transfer $t$, in the present analysis it is taken to be constant since we deal with reactions proceeding at conditions comparable
to the case of the decay $\pi_{e2\gamma}$ ($t\sim m^2_{\pi}$)~\cite{Chen2011}. 

Consider matter containing photons and pions. If thermal equilibrium takes place at temperature $T$ and the medium is transparent to the outgoing neutrinos (i.e. there is no Pauli blocking for the final state particles), the energy loss rate per unit volume due to emission of neutrinos of flavor $l$ (the emissivity) through the reaction~(\ref{fi0}) is given by

\begin{equation}
Q_{\nu\bar\nu}=\frac{2}{(2\pi)^6}\int\frac{d^3{\bf k_{\gamma}}}{\left[\exp{(\omega_{\gamma}/T)-1}\right]}\frac{d^3{\bf k_{\pi}}}{\left[\exp{(\omega_{\pi}/T)-1}\right]}(\omega_{\gamma}+\omega_{\pi})\sigma_{\pi}v_r, \label{emmis}
\end{equation}

where $\omega_{\gamma}$ and $\omega_{\pi}$ are the photon and pion energies, respectively, ${\bf k_{\gamma}}$ and ${\bf k_{\pi}}$ are their three-momenta, $v_r$ is the relative velocity

\begin{equation}
v_r=\frac{\omega_{\gamma}\omega_{\pi}-{\bf k_{\gamma}}\cdot{\bf k_{\pi}}}{\omega_{\gamma}\omega_{\pi}}.
\label{velocity}
\end{equation}

The pion vector form factor, $F_V$, is related via the Conserved Vector Current  hypothesis~(CVC) to the $\pi^0\rightarrow\gamma\gamma$ decay width $\Gamma_{\pi^0\rightarrow{\gamma\gamma}}$ by~\cite{vfactor1,vfactor2}

\begin{equation}
|F_V|^2=\frac{2\Gamma_{\pi^0\rightarrow{\gamma\gamma}}}{\alpha^2\pi m_\pi}.
\label{pion_number}
\end{equation}

Let us rewrite the cross section (\ref{cross_sect_tot}) by invoking the CVC:

\begin{equation}
\sigma_{\pi}=\frac{G^2_{F}}{6\alpha\pi m^3_\pi}\Gamma_{\pi^0\rightarrow{\gamma\gamma}}\left(m^2_{\pi}+2\omega_{\gamma}\omega_{\pi}v_r\right)\omega_{\gamma}\omega_{\pi}v_r, 
\label{crosspi0}
\end{equation}

Note that $s=m^2_{\pi}+2\omega_{\gamma}\omega_{\pi}v_r$.

Calculations of the momentum space integrals in (\ref{emmis}) taking into account (\ref{velocity}) and (\ref{crosspi0}) give

\begin{equation}
Q_{\nu\bar\nu}=\frac{G_F^2}{3780\alpha\pi^5m_{\pi}^3}\Gamma_{\pi^0\rightarrow{\gamma\gamma}}T^4\,{\cal I}(m_{\pi},T), \label{emmis2}
\end{equation}

where 

\begin{eqnarray}
{\cal I}(m_{\pi},T)=\int^\infty_{m_{\pi}}d\omega_{\pi}\frac{\sqrt{\omega_{\pi}^2-m^2_{\pi}}}{\exp(\omega_{\pi}/T)-1}\left[2520\zeta(5)T(12\omega_{\pi}^4-2m_{\pi}^2\omega_{\pi}^2-m_{\pi}^4)-\right.\nonumber\\\left.-\pi^4\omega_{\pi}(80\pi^2m_{\pi}^2T^2-4\omega_{\pi}^2(7m_{\pi}^2+40\pi^2T^2)+7m_{\pi}^4)\right],\label{intergral}
\end{eqnarray}

\begin{figure}
\centering
\includegraphics{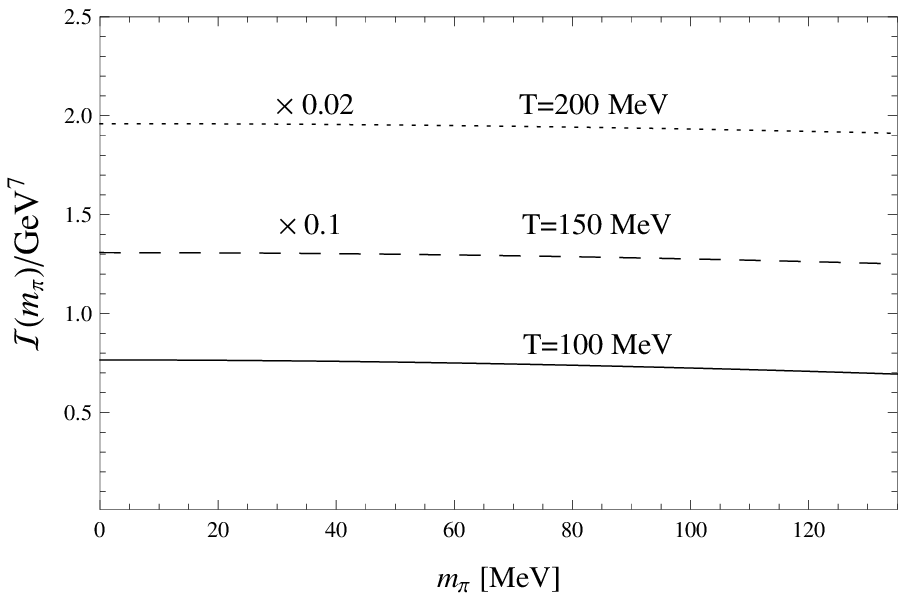}
\caption{}
\label{fig2}
\end{figure}

$\zeta(x)$ is the Riemann zeta-function ($\zeta(5)=1.037$). Note that the integral~(\ref{intergral}) very weakly depends on the pion mass in the range considered in this Letter (see Fig.~\ref{fig2}). The change of ${\cal I}(m_{\pi},T)$ in the interval $0\leq m_{\pi}\leq135$ MeV constitutes only few percent, so that one can take the integral with good accuracy mass independent. This approximation becomes even much better for small masses and higher temperatures.

It appears that the proposed process $\gamma\pi^0\rightarrow\nu_l\bar\nu_{l}$ mimics the reactions $\pi^0\rightarrow\nu_l\bar\nu_{l}$ and $\gamma\gamma\rightarrow\pi^0\rightarrow\nu_l\bar\nu_{l}$~\cite{pion_cond2} and should also be considered in studies as~\cite{pion_cond2}. Moreover, there are conditions under which the process $\gamma\pi^0\rightarrow\nu_l\bar\nu_{l}$ dominates over $\pi^0\rightarrow\nu_l\bar\nu_{l}$ and $\gamma\gamma\rightarrow\pi^0\rightarrow\nu_l\bar\nu_{l}$. The matter is that the rates of the latter two reactions are proportional to the width of the decay $\pi^0\rightarrow\nu_l\bar\nu_l$~\cite{pion_cond2} whose experimental value is not yet convincingly determined and varies in the very wide range -- from $\sim$$10^{-13}$ eV~\cite{astrophys} to $\sim$$10^{-6}$ eV~\cite{pdg}. This fact is inevitably reflected in the results of calculations of the related neutrino emissivities strongly depending on $\Gamma(\pi^0\rightarrow\nu_l\bar\nu_{l})$ adopted. In addition, the neutrino emission through $\gamma\gamma\rightarrow\pi^0\rightarrow\nu_l\bar\nu_{l}$ varies in the pion pole as $1/\Gamma_{\pi^0\rightarrow\gamma\gamma}$ and is therefore sensitive to the change of the $\pi^0\rightarrow\gamma\gamma$  decay width. Meanwhile, $\Gamma_{\pi^0\rightarrow\gamma\gamma}$ in the medium may be orders of magnitude larger than in vacuum~\cite{pion_cond2,raffelt} causing thus a suppression of the corresponding neutrino emission. In contrast to this situation, the growth of $\Gamma_{\pi^0\rightarrow\gamma\gamma}$  leads to a significant enhancement of the energy loss through $\gamma\pi^0\rightarrow\nu_l\bar\nu_{l}$ (see section~{\ref{sec4}}). 

\begin{figure}
\centering
\includegraphics{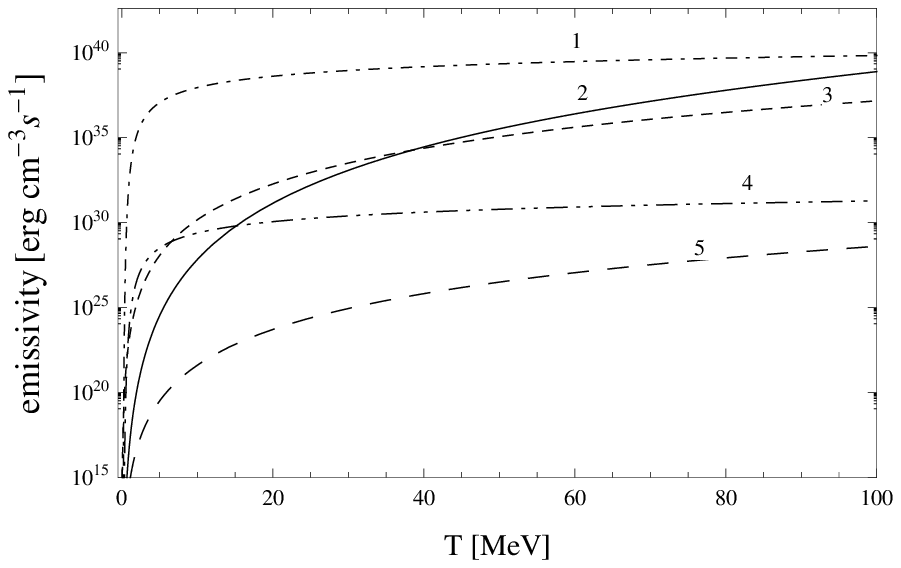}
\caption{}
\label{fig3}
\end{figure}

The temperature dependences of the emissivities are depicted in Fig.~\ref{fig3}. The calculations are performed assuming the vacuum values of $m_{\pi}$ and $\Gamma_{\pi^0\rightarrow{\gamma\gamma}}$. One can see that $Q_{\nu\bar\nu}$ largely dominates over the neutrino emissivities  through the competing process $\pi^0\rightarrow\nu_l\bar\nu_{l}$ for $T\gtrsim15$ MeV and through  $\gamma\gamma\rightarrow\pi^0\rightarrow\nu_l\bar\nu_{l}$~\cite{pion_cond2} for all considered temperatures provided the astrophysical limit on $\Gamma(\pi^0\rightarrow\nu_l\bar\nu_l)$~\cite{astrophys} is used. Even if one assumes that $\Gamma(\pi^0\rightarrow\nu_l\bar\nu_l)$ is equal to its upper experimental limit~\cite{pdg}, the contribution of $\gamma\pi^0\rightarrow\nu_l\bar\nu_{l}$ to the emissivity remains to be dominant over that of the pion pole mechanism $\gamma\gamma\rightarrow\pi^0\rightarrow\nu_l\bar\nu_{l}$ for temperatures $T\gtrsim35$ MeV. Regardless of this, the analyzed processes have small impact compared to the Urca and modified Urca processes over most of the temperature range considered.

\section{Single neutrino photoproduction on charged pNGB}
Photons are also able to produce single neutrinos on charged pNGB:

\begin{equation}
\gamma+\pi^+\rightarrow l^++\nu_l,
\label{charged}
\end{equation}

where $l=e, \mu$.

The Feynman diagrams contributing to (\ref{charged}) are shown in Fig.~\ref{fig4}.

The corresponding matrix element is~\cite{matrix_cite1,matrix_cite2}

\begin{equation}
{\cal M}={\cal M}_{a}+{\cal M}_{b}+{\cal M}_{c}
\label{matrix_sum}
\end{equation}

with

\begin{equation}
{\cal M}_a+{\cal M}_{b}=-ie\frac{G_F}{\sqrt{2}}V_{ud}f_{\pi}m_{l}\varepsilon_{\mu}\bar u(p_{\nu})(1+\gamma_5)\left(\frac{p^{\mu}_{\pi}}{p_{\pi}\cdot q}-\frac{2p^{\mu}_l-\not \!{q}\gamma^{\mu}}{2p_{l}\cdot q}\right)v(p_{l}),
\label{matrix_ab}
\end{equation}

\begin{eqnarray}
{\cal M}_c=ie{\frac{G_{F}}{\sqrt{2}}}V_{ud}\varepsilon_{\mu }
\bar{u}(p_{\nu})\gamma _{\alpha }(1-\gamma
_{5})v(p_{l})\times\hskip 4cm\nonumber\\\times
\left[\frac{F_{A}}{m_{\pi}}(-g^{\mu \alpha }p_{\pi}\cdot
q+p_{\pi}^{\mu }q^{\alpha })+i\frac{F_{V}}{m_{\pi}}\epsilon ^{\mu \alpha \beta
\lambda }q_{\beta }p_{\pi \lambda }\right], \label{ampl_decay}
\end{eqnarray}

where $V_{ud}$ is the Cabibbo--Kobayashi--Maskawa matrix element, $f_{\pi}$ is the pion decay constant, $F_{A}$ is
the pion axial-vector form factor.

Let us restrict ourselves to a consideration of the single electron neutrino photoproduction: $\gamma\pi^+\rightarrow e^+\nu_{e}$. In  this case, unless the in-medium mass of the pion becomes very small,~$m_{\pi}\approx m_{e}$, one can safely neglect the contributions of the diagrams~(a) and~(b) for they are helicity suppressed being proportional to $m_e$ exactly as in the decay~$\pi^+\rightarrow e^+\nu_e$~(see eq.~(\ref{matrix_ab})).

Then, keeping only the contribution of the diagram~(c), which is free of the helicity suppression,  and squaring~(\ref{matrix_sum}) yields

\begin{equation}
\sum_{\text{spins}}|{{\cal M}}_c|^2=\frac{\alpha \pi G_F^2}{m_{\pi}^2}|V_{ud}|^2s\left\{t^2|F_V+F_A|^2+u^2|F_V-F_A|^2\right\}\label{matrix_c}
\end{equation}

so that the corresponding cross section reads

\begin{equation}
\sigma^c_{\pi}=\frac{\alpha G^2_{F}}{24m^2_\pi}|V_{ud}|^2\left(|F_{V}|^2+|F_{A}|^2\right)s^2\left(1-\frac{m^2_{\pi}}{s}\right). \label{cross_sect_tot2}
\end{equation}

Making the same assumptions as in section~\ref{sec0} one arrives at the following relation connecting $Q_{\nu\bar\nu}$ with the neutrino emissivity through the process~$\gamma\pi^+\rightarrow e^+\nu_e$ (denoted by $Q_{\nu}$): 

\begin{equation}
\frac{Q_{\nu}}{Q_{\nu\bar\nu}}=\frac{1}{2}|V_{ud}|^2\left(1+\frac{|F_A|^2}{|F_V|^2}\right).\label{ratio}
\end{equation}

In~(\ref{ratio}) the factor 1/2 takes into account the fact that the neutrino in the reaction $\gamma\pi^+\rightarrow e^+\nu_e$ carries away only a half of the total energy. Numerically, at the vacuum values $|V_{ud}|^2=0.9482$, $F_V=0.0272$ and $F_A=0.0112$~\cite{Chen2011},~(\ref{ratio}) gives $Q_{\nu}/Q_{\nu\bar\nu}=0.5544$.

Since the absolute square of the matrix element for the process~(\ref{charged}) including the contributions of the diagrams (a) and (b) may be useful for similar calculations, its full form is given in  the appendix.

\section{Neutrino emissivity in a hot and dense medium\label{sec4}}
In the previous sections it was assumed that the width of the pion as well as its mass do not depend on temperature. Meanwhile, these particle properties are expected to be modified under extreme conditions which can be encountered inside some stars or fabricated in ultrarelativistic heavy-ion collisions. It is therefore interesting to investigate the neutrino photoproduction on PNGb in a hot and dense medium. 

There is a set of models based on the SU(2) Nambu--Jona-Lasinio (NJL) model~\cite{jona1,jona2} predicting an enhancement of the neutral pion decay width at the so-called Mott temperature at which the pion dissociates into quark--antiquark pairs~\cite{hashimoto,klevansky,blaschke,caldas}. Since the neutrino emissivity found in this Letter is proportional to $\Gamma_{\pi^0\rightarrow{\gamma\gamma}}$ (see (\ref{emmis2})), one may expect that it will also grow at some critical temperature. 

Let us implement the thermal effects into our calculations through the pion decay constant and the pion mass following the model given in~\cite{caldas}. Within a renormalized version of the NJL model~\cite{mota},  in the chiral limit, one arrives at the well known result~\cite{fp1,fp2,fp3}

\begin{equation}
\Gamma_{\pi^0\rightarrow{\gamma\gamma}}(T)=\frac{m^3_{\pi}(T)}{64\pi}\left(\frac{N_Ce^2}{12\pi^2f_{\pi}(T)}\right)^2,\label{g_fp}
\end{equation}

where $N_C$ is the number of colors. Hereafter $N_C$ is taken to be equal to~3.

A substitution of~(\ref{g_fp}) into~(\ref{emmis2}) yields

\begin{equation}
Q_{\nu\bar\nu}=\frac{\alpha G_F^2}{241920\pi^8}\frac{T^4}{f_{\pi}^2(T)}\,{\cal I}(T). \label{emmis22}
\end{equation}

It is notable that the emissivity turns out to be independent on $m_{\pi}(T)$ in the considered energy range. This property follows from the mass dependence of the integral (\ref{intergral}) (see Fig.~\ref{fig2}) and is very convenient since in such a case one should not study the influence of the extreme conditions on the pion mass and take it into account. The thermal effects are completely determined by the pion decay constant. That is why in~(\ref{emmis22}) we write ${\cal I}(T)$ instead of ${\cal I}(m_{\pi}(T),T)$.

Now it is obvious from~(\ref{emmis22}) that models in which the pion decay constant drops down with increasing temperature predict an enhancement of the neutrino emissivity through the reaction $\gamma\pi^0\rightarrow \nu_l\bar\nu_{l}$. For example, the model described in~\cite{caldas} accommodates just the case. The thermal behavior of $f_{\pi}$ taken from~\cite{caldas} is presented in Fig.~\ref{fig5}. Using this parameterization of $f_{\pi}$ in~(\ref{emmis22}), one obtains the temperature dependence of the emissivity of neutrinos of flavor $l$ in a hot and dense medium shown in Fig.~\ref{fig6}. As expected, a fast growth of the emissivity by orders of magnitude near the critical temperature is clearly observed.

The extreme conditions can be created in ultrarelativistic heavy-ion collisions. Dileptons can serve as crucial probes in the study of matter under such conditions~\cite{shuryak1, shuryak2}. The basic quantity for connecting our calculations to measurements in heavy-ion collisions is the number of neutrino pairs emitted per unit space-time~\cite{prd_last}:

\begin{equation}
\frac{dN^{\nu}}{d^4x}=\frac{2}{(2\pi)^6}\int\frac{d^3{\bf k_{\gamma}}}{\left[\exp{(\omega_{\gamma}/T)-1}\right]}\frac{d^3{\bf k_{\pi}}}{\left[\exp{(\omega_{\pi}/T)-1}\right]}\sigma_{\pi}v_r, \label{emmis_nu}
\end{equation}

Evaluations of the integrals are quite similar to the case of~(\ref{emmis}). One can find that

\begin{equation}
\frac{dN^{\nu}}{d^4x}=\frac{3\alpha G_F^2\zeta(5)}{\pi^8}\frac{T^{10}}{f_{\pi}^2(T)}. \label{emmis_nu2}
\end{equation}
  
The volume element is $d^4x=d^2x_Tdytdt$, where $t$ is the proper time and $y$ is the rapidity of the "fireball" emitting the neutrinos. In central collisions of equal mass ions $d^2x_T=\pi R_A^2$, where $R_A$ is the ion radius. Then

\begin{equation}
\frac{dN^{\nu}}{dy}=\pi R_A^2\int_{t_0}^{t_1}dt\,t\frac{dN^{\nu}}{d^4x}(T(t)). \label{emmis_nu3}
\end{equation}
Here $t_1-t_0$ is the time interval during which the state of matter exists. One needs to know how the matter evolves in time (in other words, one needs to know the time dependence of temperature $T(t)$). Let us use the model described in detail in~\cite{prd_last}. It is convenient to change the variable of integration in~(\ref{emmis_nu3})  from $t$ to $T$:

\begin{equation}
\frac{dN^{\nu}}{dy}=3\pi R_A^2T_i^6t_0^2\int_{T_f}^{T_c}\frac{dT}{T^7}\frac{dN^{\nu}}{d^4x}, \label{emmis_nu4}
\end{equation}

where $T_i$ is the initial temperature of the thermalized quark-gluon matter, $T_c$ is the temperature of the transition from the quark-gluon phase to pure hadron phase (the critical temperature), $T_f$ is the temperature at which the hadron phase breaks up into free hadrons. For ultrarelativistic Pb-Pb collisions at the LHC energies, typically $T_i\approx 300$ MeV, $t_0\approx12$~fm, $R_A\approx7$~fm, $T_f\approx140$~MeV. The evolution of $f_{\pi}$ adopted in this Letter assumes $T_c\approx 200$ MeV.

Apart from the pure hadronic phase, there may appear the mixed phase for which the neutrino emission rate is given by~\cite{prd_last} 

\begin{equation}
\frac{dN^{\nu}_{\text{mixed}}}{dy}=\frac{\pi R_A^2}{2}\left(\frac{T_i}{T_c}\right)^6t_0(t_0-t_i)\left.\frac{dN^{\nu}}{d^4x}\right|_{T=T_c}, \label{emmis_nu5}
\end{equation}

where $t_i$ is the thermalization time ($t_i\approx1$ fm). 

Our calculations of the neutrino emission rates using~(\ref{emmis_nu4}) and~(\ref{emmis_nu5}) for ultrarelativistic Pb-Pb collisions at $y=0$ are shown in Fig.~\ref{fig7}. One can see that the in-medium modification of the pion decay constant leads to an orders of magnitude enhancement of the neutrino emission as temperature approaches its critical value. The neutrinos can be directly detected. The main contribution to the background in the region $y=0$ arises from the leptonic decays of $\pi^{\pm}$ and $K^{\pm}$ mesons at rest which produce neutrinos with fixed energies: $E_{\nu}=70$ MeV for $\pi^{\pm}\rightarrow e^{\pm}\overset{(-)}{\nu_e}$; $E_{\nu}=30$ MeV for $\pi^{\pm}\rightarrow \mu^{\pm}\overset{(-)}{\nu_{\mu}}$; $E_{\nu}=247$ MeV for $K^{\pm}\rightarrow e^{\pm}\overset{(-)}{\nu_e}$ and $E_{\nu}=236$ MeV for $K^{\pm}\rightarrow \mu^{\pm}\overset{(-)}{\nu_{\mu}}$. These background events can be well separated out from the continuous spectrum of the neutrinos coming from the reaction $\gamma\pi^0\rightarrow\nu_l\bar\nu_l$ and the latter may provide thus a distinctive signature of appearance of hot and dense matter. Apart from a direct observation of the neutrinos, the considered process  can in principle manifest itself in the form of missing energy in  experiments studying the extensive air showers initiated by ultra-high energy cosmic ray nuclei in Earth's atmosphere provided the fall of $f_{\pi}(T)$ with increasing temperature is rather dramatic than the one predicted by the model adopted in this Letter. Simultaneously this will generate an additional flux of atmospheric neutrinos. 

\section{Conclusions}
Photoproduction of neutrino--antineutrino pairs and single neutrinos in the reactions $\gamma\pi^0\rightarrow \nu_l\bar\nu_{l}$, $\gamma\pi^+\rightarrow l^+\nu_{l}$ is studied within the Standard Model. The corresponding cross sections are found analytically. 

The energy loss due to neutrino emission in a thermal plasma of photons and pions is calculated. It is shown that the obtained neutrino emissivities may be significantly enhanced in hot and dense matter due to in-medium modification of the pion decay constant. This process has fascinating phenomenological consequences. 
In ultrarelativistic heavy-ion collisions, for example, this will yield a background of directly produced neutrinos with an evaporation-like spectrum in the center-of-mass frame. It is noticeable that its emergence in interactions of the primary ultra-high energy cosmic ray nuclei with the atomic nuclei of Earth's atmosphere or Moon rock will generate an additional flux of neutrinos. This phenomenon may manifest itself in the form of missing energy in the cosmic ray energy spectrum because the experimentally unregistered neutrinos will carry away a fraction of the total collision energy. In this connection, it is interesting to speculate on the origin of the knee in the energy spectrum of the primary cosmic rays~\cite{petrukhin}.  

This mechanism might also play an important role in neutrino production in the early universe as well as in thermal evolution of astrophysical objects containing pseudoscalar excitations such as supernovae and compact stars. 

The analysis of this Letter is closely related to the problem of pion stability in a hot medium. The reaction $\gamma\pi^0\rightarrow \nu_l\bar\nu_{l}$ mimics the decay $\pi^0\rightarrow\nu_l\bar\nu_l$ but in difference from the latter can proceed even in the pion rest frame at $m_{\nu}=0$ and with considerable probability. In other words, the corresponding cross section does not vanish in the limit of massless neutrinos and observation  of neutrino--antineutrino pairs does not therefore require the assumption of Lorentz invariance violation. 

The results concerning the neutrino photoproduction on charged pions directly apply to the case of the charged kaon target. One has just to perform replacements of the appropriate parameters in the formulae (namely $m_{\pi}\rightarrow m_{K}$, $V_{ud}\rightarrow V_{us}$, $f_{\pi}\rightarrow f_{K}$, $F_{V,A}\rightarrow F_{V,A}^K$). In addition, 
the presented calculations will be exactly the same for $\gamma(\pi^-,K^-)\rightarrow l^-\bar\nu_{l}$ if CP is conserved.

\vskip 0.5cm
{\bf Acknowledgements}
\vskip 0.5cm
This work was supported in part by the Russian Foundation for Basic Research (grant 11-02-12043), by the Program for Basic Research of the Presidium of the Russian Academy of Sciences "Fundamental Properties of Matter and Astrophysics" and by the Federal Target Program  of the Ministry of Education and Science of Russian Federation "Research and Development in Top Priority Spheres of Russian Scientific and Technological Complex for 2007-2013" (contract No. 16.518.11.7072).

\begin{figure}
\centering
\resizebox{0.9\textwidth}{!}{%
\includegraphics{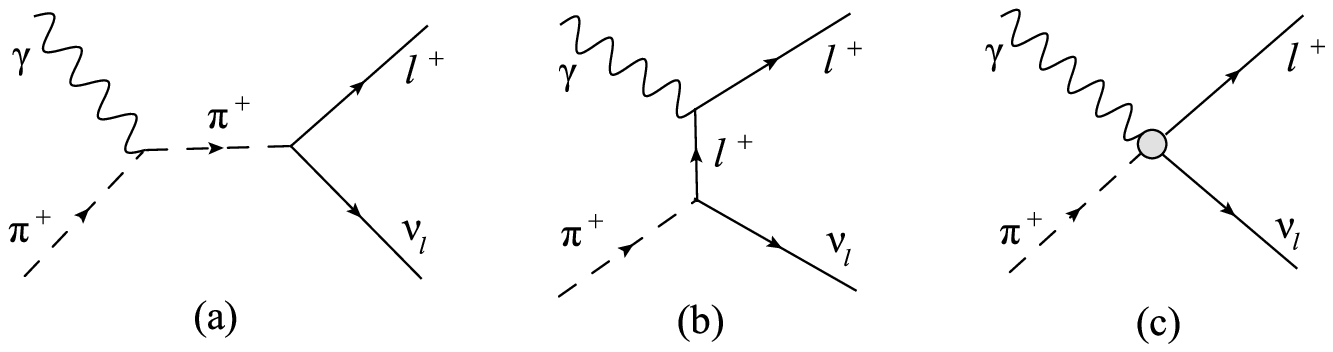}
}\caption{}
\label{fig4}
\end{figure}

\begin{figure}
\centering
\includegraphics{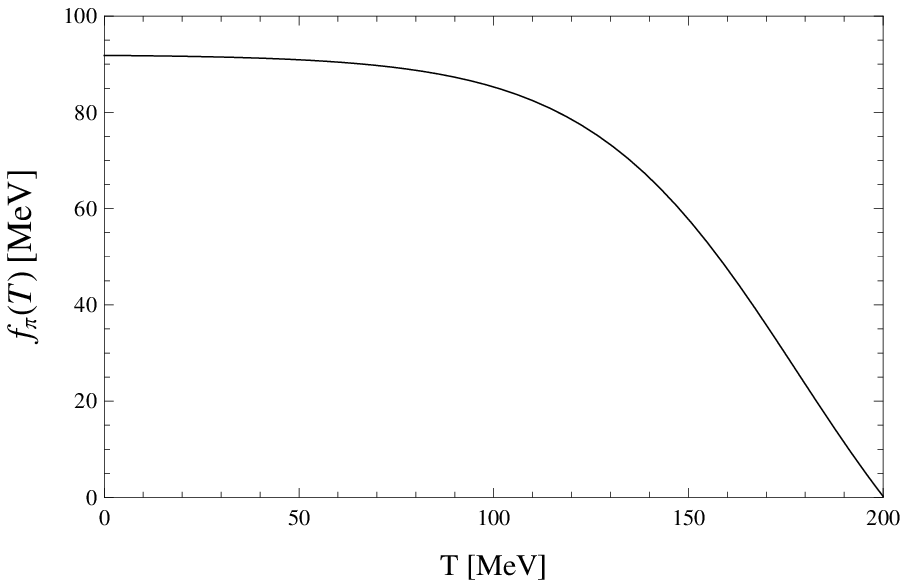}
\caption{}
\label{fig5}
\end{figure}

\begin{figure}
\centering
\includegraphics{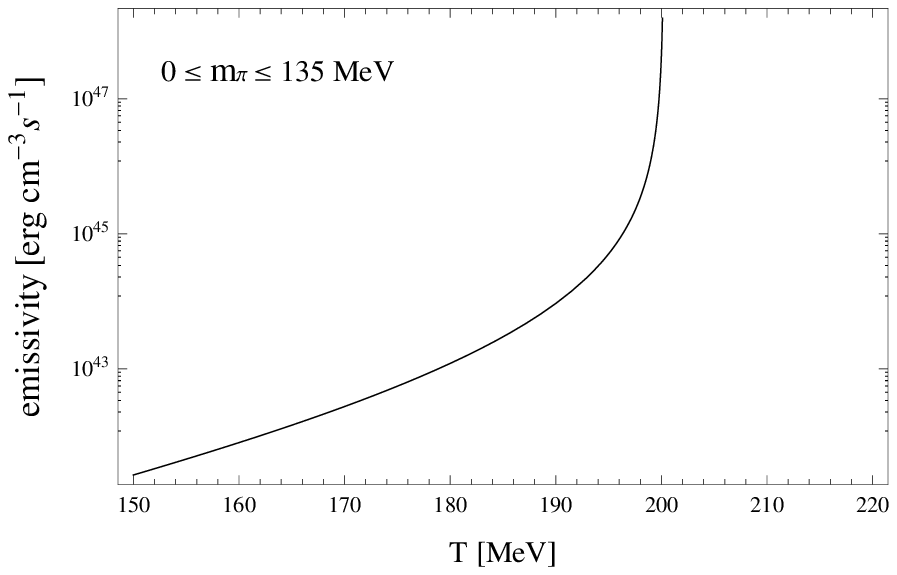}
\caption{}
\label{fig6}
\end{figure}

\begin{figure}
\centering
\includegraphics{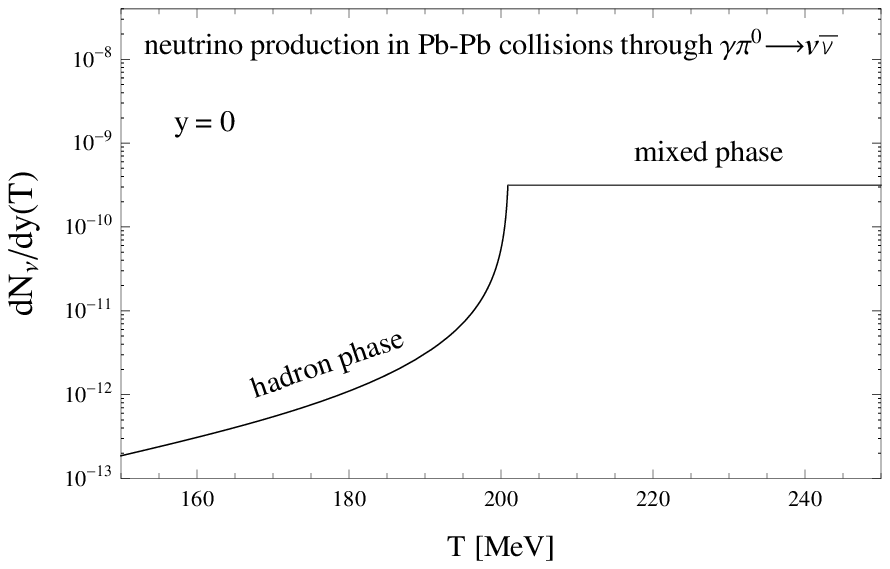}
\caption{}
\label{fig7}
\end{figure}

\newpage
{\bf Appendix}
\vskip 0.5cm
The absolute square of the matrix element for the process (\ref{charged}) represented by the Feynman diagrams in Fig.~\ref{fig4}: 

\begin{eqnarray}
\sum_{\text{spins}}|{\cal M}_{a}+{\cal M}_{b}+{\cal M}_{c}|^2=\alpha \pi G_F^2|V_{ud}|^2\times\hskip 7cm\nonumber\\\times\left\{4f_{\pi}m_{l}^2u\left(\frac{f_{\pi}(2m_l^4m_{\pi}^2-m^2_l(2m_{\pi}^4+(s-m_{\pi}^2)^2+2st)+t(m_{\pi}^4+s^2))}{(t-m_l^2)^2(s-m_{\pi}^2)^2}-\right.\right.\nonumber\\\left.\left.-\frac{Re[(F_V+F_A)^*](m_l^2m_{\pi}^2-st)-Re[(F_V-F_A)^*]((m_{\pi}^2-m_l^2)(s-m_{\pi}^2)+su)}{m_{\pi}(t-m_l^2)(s-m_{\pi}^2)}\right)+\right.\nonumber\\\left.+\frac{1}{m_{\pi}^2}\left(|F_V+F_A|^2(m_l^2-t)(m_l^2m_{\pi}^2-st)-|F_V-F_A|^2u(m_{l}^2(s-m_{\pi}^2)-su)\right)\right\}.\label{matrix_tot}
\end{eqnarray}

One can see that~(\ref{matrix_tot}) in the limit $m_l=0$ is reduced to~(\ref{matrix_c}).


\newpage

{\bf Figure Captions}
\vskip 0.5 cm 

{\bf Fig. 1:} Feynman diagram for the process $\gamma\pi^0\rightarrow\nu_{l}\bar\nu_{l}$.

\vskip 0.5 cm 

{\bf Fig. 2:} The integral (\ref{intergral}) as a function of the pion mass at three fixed values of temperature.

\vskip 0.5 cm

{\bf Fig. 3:} The temperature dependences of the energy loss rates due to emission of neutrinos of flavor $l$ through the following reactions: (1) $\pi^0\rightarrow\nu_l\bar\nu_{l}$ assuming the experimental upper limit for $\Gamma(\pi^0\rightarrow\nu_l\bar\nu_l)$~\cite{pion_cond2}; (2) $\gamma\pi^0\rightarrow\nu_l\bar\nu_{l}$; (3) $\gamma\gamma\rightarrow\pi^0\rightarrow\nu_l\bar\nu_{l}$ assuming the experimental upper limit for  $\Gamma(\pi^0\rightarrow\nu_l\bar\nu_l)$~\cite{pion_cond2}; (4) $\pi^0\rightarrow\nu_l\bar\nu_{l}$ assuming the astrophysical limit for $\Gamma(\pi^0\rightarrow\nu_l\bar\nu_l)$ from~\cite{astrophys}; (5) $\gamma\gamma\rightarrow\pi^0\rightarrow\nu_l\bar\nu_{l}$ assuming the astrophysical limit for  $\Gamma(\pi^0\rightarrow\nu_l\bar\nu_l)$ from~\cite{astrophys}. All curves are obtained assuming the vacuum values of $m_{\pi}$ and $\Gamma_{\pi^0\rightarrow{\gamma\gamma}}$.

\vskip 0.5 cm

{\bf Fig. 4:} Feynman diagrams for the process $\gamma\pi^+\rightarrow l^+\nu_{l}$ ($l=e, \mu$).

\vskip 0.5 cm

{\bf Fig. 5:} The pion decay constant as a function of temperature taken from~\cite{caldas}.

\vskip 0.5 cm

{\bf Fig. 6:} The emissivity of neutrinos of flavor $l$ through the reaction~$\gamma\pi^0\rightarrow\nu_l\bar\nu_{l}$ in a hot and dense medium as a function of temperature.

\vskip 0.5 cm

{\bf Fig. 7:} The neutrino emission rate in ultrarelativistic Pb-Pb collisions through the reaction~$\gamma\pi^0\rightarrow\nu_l\bar\nu_{l}$ as a function of temperature at $y=0$. Two possibilities are taken into account: the pure hadronic phase and mixed phase. 
\vskip 0.5 cm


\newpage

\end{document}